\documentclass[aps,prd,showpacs,amssymb,floatfix,twocolumn]{revtex4}
\usepackage{graphicx}
\usepackage[sort&compress]{natbib}
\begin{document}
%%%   New Definitions
\newcommand{\eg}{{\it e.g.}}
\newcommand{\etal}{{\it et. al.}}
\newcommand{\ie}{{\it i.e.}}
\newcommand{\be}{\begin{equation}}
\newcommand{\dd}{\displaystyle}
\newcommand{\ee}{\end{equation}}
\newcommand{\bea}{\begin{eqnarray}}
\newcommand{\eea}{\end{eqnarray}}
\newcommand{\bef}{\begin{figure}}
\newcommand{\eef}{\end{figure}}
\newcommand{\bce}{\begin{center}}
\newcommand{\ece}{\end{center}}
\def\lsim{\mathrel{\rlap{\lower4pt\hbox{\hskip1pt$\sim$}}
    \raise1pt\hbox{$<$}}}         %less than or approx. symbol
\def\gsim{\mathrel{\rlap{\lower4pt\hbox{\hskip1pt$\sim$}}
    \raise1pt\hbox{$>$}}}         %greater than or approx. symbol

\title{Astrophysical constraints on the confining models : the
Field Correlator Method}
\author{M. Baldo, G. F. Burgio, P. Castorina, S. Plumari and  D. Zappal\`a}
\affiliation{INFN Sezione di Catania and Dipartimento di Fisica e Astronomia,
Universit\`a di Catania, Via Santa Sofia 64,
I-95123 Catania, Italia}

\date{\today}
\begin{abstract}
We explore the relevance of confinement in quark matter models for the possible quark core
of neutron stars. For
the quark phase, we adopt the equation of state (EoS) derived with the Field Correlator
Method, extended to the
zero temperature limit. For the hadronic phase, we use the microscopic Brueckner-Hartree-
Fock many-body theory.
We find that the currently adopted value of the gluon condensate $G_2 \simeq 0.006-0.007~\rm
{GeV^4}$, which gives a
critical temperature $T_c \simeq~170~ \rm MeV$, produces maximum masses which are only
marginally consistent
with the observational limit, while larger masses are possible if the gluon condensate is
increased.

\end{abstract}

\maketitle

------------------------------------------------------------

\section{Introduction}

QCD at finite temperature and density is the essential theoretical tool to
describe various interesting phenomenological sectors from relativistic heavy
ion collisions to the inner structure of neutron stars.
In the  large temperature and small density region both experiments
\cite{rhic} and lattice simulations \cite{lqcd} clearly indicate that
the theory is
in a non-perturbative regime at least up to temperatures $ T \simeq 3 T_c(0)$
($T_c(0)$ is the deconfinement temperature at zero quark chemical
potential $\mu_q=0$).
However, in the opposite region of the phase diagram, i.e.  at small $T$ and
large $\mu_q$, where strong coupling effects are expected as well,
no QCD lattice simulations are available yet.

Due to the lack of lattice data, analytic approaches based on more elementary
models, such as the Nambu--Jona-Lasinio (NJL) model \cite{bubtes}, that mimic some
non-perturbative features of QCD, are mostly used in the large density
region, typical of neutron star interior. Unfortunately the NJL model cannot
be used in the other limit of zero
chemical potential and high temperature because of the lack of the gluon
degrees of freedom. This is a general
feature of many models, which cannot make predictions for both limits, i.e.
high temperature and zero chemical
potential or high chemical potential and low temperature. This is clearly a
serious drawback, since the models cannot be fully tested.
One of the few exceptions is the Field Correlator Method (FCM) \cite{phrep},
which in principle
is able to cover the full temperature-chemical potential plane.
Furthermore the method contains {\it ab initio} the property of
confinement, which is expected to play a role, at variance with other models
like, e.g., the NJL model.

The study of the properties of neutron stars (NS) concerns the large density
(and low temperature) region of the  phase diagram and, in particular,
it requires the QCD non-pertubative Equation of State at small $T$ and
large $\mu_q$. The comparison of
the quark matter EoS with the nuclear matter one is the main point to
understand if a core of pure quark matter can  exist in NS.
This possibility has been
addressed and extensively discussed in the literature \cite{gle,ozel,alf}.
In the NJL model, where the phase transition corresponds to chiral symmetry
restoration, the quark onset at the center of the NS, as the mass increases,
marks an instability of the star, i.e. the NS collapses to a black hole at the
transition point since the quark EoS is unable to sustain the
increasing central  pressure due to gravity. Indeed, at the maximum mass the
mass-radius relation is characterized by a cusp \cite{inst}. On
the contrary, for the quark EoS based on the MIT bag model, it is possible to
find a range of the various  parameters which corresponds to a stable NS.
It must be noted that stability is also found in other approaches that
explicitly take into account the dynamics of confinement, such as the
dielectric model \cite{col} or a
modification of the NJL with an {\it ad hoc} confining potential \cite{thom}.
A modified NJL with the explicit inclusion of colour superconductivity
and isoscalar vector meson coupling \cite{bla} produces stable NS as well.
This shows how the presence of
quark matter in the interior of NS depends on the adopted quark matter model.

The intriguing relation between
the stability of NS with a quark matter core and confinement has already been
addressed in \cite{noi}, and in
this paper we elaborate further on this idea by resorting to the EoS of the
quark matter at finite temperature
and density, obtained in the non perturbative framework of the Field Correlator
Method (for a review see \cite{phrep}), which gives a natural explanation and
treatment of the dynamics of confinement in terms of Color
Electric (CE) and Color Magnetic (CM) field correlators. In this way the FCM
will be tested by comparing the results for the neutron star masses with the
existing phenomenology, which turns out to be a strong constraint
on the parameters used in the model.

It will be shown that this approach, unlike the non-confining NJL model,
admits stable NS with gravitational masses slightly larger than 1.44 $M_\odot$,
and this application of the FCM to the study of NS, which has not been
considered before, provides definite numerical indications on some relevant
physical quantities, as the gluon condensate, to be compared to
the ones extracted from the determination of the
critical temperature of the deconfinement phase
transition. This shows the relevance
of the comparison of the model predictions
in the high chemical potential region
with the astrophysical phenomenology,
which is one of the main purposes of the present paper.

In the next section the FCM at finite temperature and density is briefly
recalled, while Sec.III contains some details of the EoS for the hadronic
phase. Our analysis of stability of the NS
is presented in Sec.IV and finally Sec.V is devoted to the conclusions.

\section{Quark Matter: EOS in the Field Correlator Method}

A systematic method to treat non perturbative effects in QCD is by gauge invariant field
correlators \cite{phrep}. The approach based on the FCM provides a natural treatment
of the dynamics of confinement (and of the deconfinement transition) in terms of the
CE  ($D^E$ and $D_1^E$)  and CM ($D^H$ and $D_1^H$) Gaussian
(i.e. quadratic in the tensor $F_{\mu\nu}^a$) correlators.
$D^C$ and $D_1^C$ are related to the simplest non trivial 2-point correlators
for the CE and CM fields by:
\bea
g^2 < Tr_f [C_i(x) \Phi(x,y) C_k(y) \Phi(y,x)]> =&&
\nonumber\\
\delta_{ik}[D^C(z) + D_1^C (z) + z_4^2 \frac{\partial D_1^C (z) }{\partial z^2}]
\pm z_i z_k \frac
{\partial D_1^C(z) }{\partial z^2}
&&\eea
where $z=x-y$, $C$ indicates the CE (E) field or CM (H) field (the minus sign in the
previous expression corresponds to the magnetic case) and finally
\be
\Phi= P \exp{\left [ ig \int_y^x A^\mu dz_\mu\right ]}
\ee
is the parallel transporter.

$D^C$ and $D_1^C$ have a perturbative contribution which is responsible of their
singular behavior at $z\sim 0$ ($ D \simeq z^{-4}$ for $z\to 0$), but also a
non-perturbative part  which is normalized to the gluon condensate \cite{phrep}.
$D^E$ contributes to the  standard string tension and is directly related to confinement,
so that its vanishing above the critical temperature implies deconfinement.

The FCM has been extended to finite temperature $T$ and chemical potential $\mu_q$
 in order to describe the deconfinement phase transition
\cite{sim1,sim3,sim4,sim22,sim5,sim6}. In particular, at $\mu_q=0$ the analytical results in the gaussian
approximation, valid for small vacuum correlation lengths,  are in reasonable agreement with lattice data
\cite{sim1,sim4,sim22}. The extension in ref. \cite{sim4} of the FCM to finite values of the chemical potential,
allows to obtain a simple expression of the Equation of State of the quark-gluon matter in the relevant range of
baryon density. The comparison of this EOS with a realistic baryonic EOS will be the crucial point of our
investigation.

It must be noticed that the generalization  of the FCM at finite $T$ and $\mu_q$
provides an expression of the pressure of quarks and gluons where the leading
contribution is given by the interaction of single quark and gluon line with the vacuum,
called Single Line Approximation (SLA), while the pair and triple correlations
yield higher order corrections.
In the SLA, within few percent, the quark pressure, for a single flavour,
is given by \cite{sim3,sim22,sim5,sim6}
\be\label{pquark}
P_q/T^4 = \frac{1}{\pi^2}[\phi_\nu (\frac{\mu_q - V_1/2}{T}) +
\phi_\nu (-\frac{\mu_q + V_1/2} {T})]
\ee
where $\nu=m_q/T$, and
\be
\phi_\nu (a) = \int_0^\infty du \frac{u^4}{\sqrt{u^2+\nu^2}} \frac{1}{(\exp{[ \sqrt{u^2 +
\nu^2} - a]} + 1)}
\ee
and $V_1$ is the large distance static $Q \bar Q$ potential:
\be
\label{v1}
V_1 = \int_0^{1/T} d\tau(1-\tau T) \int_0^\infty d\chi \chi D_1^E(\sqrt{\chi^2 + \tau^2})
\ee

The gluon contribution to the pressure is
\be\label{pglue}
P_g/T^4 = \frac{8}{3 \pi^2} \int_0^\infty  d\chi \chi^3
\frac{1}{\exp{(\chi + \frac{9 V_1}{8T} )} - 1}
\ee

Note that the potential  $V_1$ in Eq.(\ref{v1}) does not depend on the
chemical potential and this is partially supported by lattice simulations
at small chemical potential \cite{sim22,latmuf}.
In our opinion, although we are considering  the range $T\sim 0$
(in the following calculations we fix the value $T=1$ MeV)
and large $\mu_q$, relevant for the NS, this approximation is still valid.
Indeed the non perturbative
contribution to $D_1^E(x)$ is parametrized as \cite{phrep}
\be
\label{d1nonpt}
D_1^E(x)=D_1^E(0) \exp(-|x|/\lambda)
\ee
where $\lambda$ is the correlation length ($0.34~\rm{fm}$ for full QCD)
and the normalization is fixed by the condition at $T=\mu=0$
\be
\label{norm}
D^E(0) +D_1^E(0) =\frac{\pi^2}{18} G_2
\ee
$G_2$ is the gluon condensate whose numerical value, determined by the QCD sum rules,
is known with large uncertainty \cite{gluecond}
\be
G_2=0.012\pm 0.006~ \rm{GeV^4}
\ee

According to \cite{sim22}, the critical temperature  at $\mu=0$ in the
FCM turns out to be $T\sim 170$ $\rm{MeV}$ for  $G_2\sim 0.006$ $\rm{GeV^4}$.
If confinement is dominated by non-perturbative contributions, the normalization $D_1^E(0)$ in Eq.
(\ref{d1nonpt}) can be indeed identified with the term appearing in Eq. (\ref{norm}), which  has been denoted by the same symbol. Then
from Eqs. (\ref{v1}), (\ref{d1nonpt}), (\ref{norm}), in the limit $T\to0$,
we get
\be
\label{constraint}
V_1(T=0)\leq \frac{\pi^2}{9} G_2 \lambda^3
\ee

However other choices of $V_1$ are possible, and these will be considered in the Discussion section.

Since, on general ground, we expect that the value of the gluon condensate decreases at large densities
\cite{balcasza}, the assumption that $V_1$ is $\mu$-independent, should not qualitatively modify our analysis.

\section{Hadronic Phase: EOS in the Brueckner-Bethe-Goldstone theory}

The EOS constructed for the hadronic phase at $T=0$ is based on the
non-relativistic Brueckner-Bethe-Goldstone (BBG) many-body theory \cite{tri},
which is a linked cluster
expansion of the energy per nucleon of nuclear matter, well convergent
and accurate enough in the density range relevant for neutron stars.
In this approach the essential ingredient is the two-body scattering matrix
$G$, which,
along with the single-particle potential $U$, satisfies the
self-consistent equations

\begin{eqnarray}
G(\rho;\omega)& = & v  + v \sum_{k_a k_b} {{|k_a k_b\rangle  Q  \langle k_a k_b|}
  \over {\omega - e(k_a) - e(k_b) }} G(\rho;\omega), \\
U(k;\rho) &= &\sum _{k'\leq k_F} \langle k k'|G(\rho; e(k)+e(k'))|k k'\rangle_a
\end{eqnarray}
\noindent
where $v$ is the bare nucleon-nucleon (NN) interaction, $\rho$ is the nucleon
number density, $\omega$  is the  starting energy, and
$|k_a k_b\rangle Q \langle k_a k_b|$  is  the Pauli operator.
$e(k) = e(k;\rho) = {{\hbar^2}\over {2m}}k^2 + U(k;\rho)$
is the single particle energy,
and the subscript ``{\it a}'' indicates antisymmetrization of the
matrix element. In the BHF approximation the energy per nucleon is
 \begin{eqnarray}
&&{E \over{A}}(\rho)  =
          {{3}\over{5}}{{\hbar^2~k_F^2}\over {2m}} + D_{\rm 2}\, , \\
&&D_{\rm 2} = {{1}\over{2A}}
\sum_{k,k'\leq k_F} \langle k k'|G(\rho; e(k)+e(k'))|k k'\rangle_a
\end{eqnarray}
\noindent
For the two-body interaction $v$, we choose the Argonne $v_{18}$
nucleon-nucleon potential \cite{v18}.
We have also introduced three-body forces
(TBF) among nucleons, adopting the phenomenological Urbana model \cite{uix}.
This allows to reproduce correctly the nuclear matter saturation point
$\rho_0 \approx 0.17~\mathrm{fm}^{-3}$, $E/A \approx -16$ MeV, and gives
values of incompressibility and symmetry energy at saturation compatible
with those extracted from phenomenology \cite{myers}.
Moreover, the BBG approach has been extended to the hyperonic sector
in a fully self-consistent way \cite{hypmat,hypns}, by including the
$\Sigma^-$ and $\Lambda$ hyperons.

In this paper, we adopt a conventional description of stellar matter, as
composed by neutrons, protons, and leptons in beta equilibrium \cite{bbb}.
The EoS for the beta equilibrated matter can be obtained for a
given composition, together with the chemical potentials of all species
as a function of the total baryon density. The chemical potentials
are the fundamental input for
the equations of chemical equilibrium, charge neutrality conditions, and baryon
number conservation, i.e.,
\begin{eqnarray}
\mu_n &= & \mu_p + \mu_{e^-} \\
\mu_{e^-}&=& \mu_{\mu^-} \\
x_p &=& x_{e^-} + x_{\mu^-} \\
1 &=& x_n + x_p
\end{eqnarray}
where $x_i=\rho_i/\rho$ is the nucleonic fraction of the species $i$.
The above conditions allow the unique solution of a closed system of equations,
yielding the equilibrium fractions of the nucleonic and leptonic species for
each fixed nucleon density.
Once the composition of the $\beta$-stable, charge neutral stellar matter
is known, one can calculate the equation of state, i.e., the relation between
pressure $P$ and energy density $\epsilon$ as a function of the baryon density
$\rho$. It can be easily obtained from the thermodynamical relation
\begin{eqnarray}
P &=& - \frac{dE}{dV} = P_B + P_l \label{phad1}\\
P_B &=& \rho^2 \frac{d(\epsilon_B/\rho)}{d\rho}, \;\; \;
P_l = \rho^2 \frac{d(\epsilon_l/\rho)}{d \rho}\label{phad2}
\end{eqnarray}
with $E$ the total energy and $V$ the total volume. The total nucleonic energy
density is obtained by adding the energy densities of each species
$\epsilon_i$. As far as
leptons are concerned, at those high densities electrons are a free
ultrarelativistic gas, whereas muons are relativistic. Hence their energy
densities $\epsilon_l$ are well known from textbooks \cite{shapiro}.

\section{Phase transition in beta-stable matter and neutron star structure}

We are now able to compare the pressure of the two phases, namely
the pressure in the hadronic phase given in Eqs.(\ref{phad1}), (\ref{phad2})
with the one in the quark-gluon phase which, according to \cite{sim22,sim6}, can
be written as
\be
\label{pqgp1}
P_{qg} =P_g+\sum_{j=u,d,s} P^j_{q} + \Delta \epsilon_{vac}
\ee
where $P_g$ and $P^j_{q}$ are respectively given in Eq. (\ref{pglue}) and
(\ref{pquark}), and
\be
\label{pqgp2}
\Delta \epsilon_{vac}
%=|\epsilon_{vac}^{dec}|-|\epsilon_{vac}^{conf}|
\approx - \frac{(11-\frac{2}{3}N_f)}{32} \frac{G_2}{2}
\ee
corresponds to the difference of the vacuum energy density in the two phases,
being $N_f$ the flavour number.
%$\Delta G_2 \approx \frac{1}{2} G_2$

By assuming a first order hadron-quark phase transition \cite{fodor} in
beta-stable matter, we adopt the simple Maxwell construction.
The more general Gibbs
construction \cite{gle} is still affected by many theoretical
uncertainties \cite{mixed}, and in any case the final mass-radius
relation of massive neutron stars \cite{mit} is slightly affected.

\begin{figure}[t] %..................................................
\centering
\includegraphics[width=12cm,angle=270]{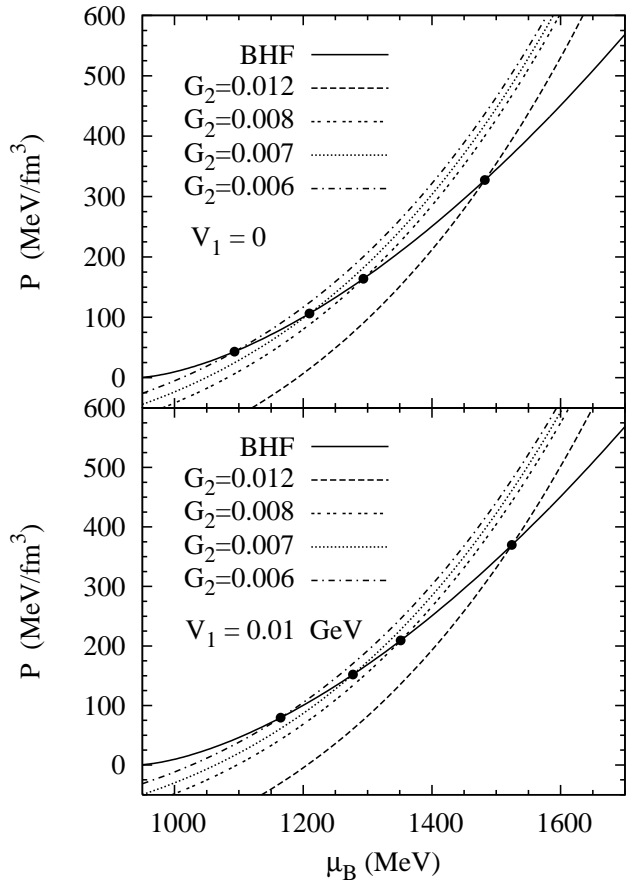}
\caption{
Pressure as a function of the baryon chemical potential.
The full line represents the BHF calculations, and the dashed ones the model
discussed in this paper with two different choices of the parameter
$V_1$, and several values of the gluon condensate $G_2$. See text for details.}
\label{f:pmu}
\end{figure} %................................................................

We impose thermal, chemical, and
mechanical equilibrium between the two phases. This implies
that the phase coexistence is determined by a crossing point in the
pressure vs. chemical potential plot, as shown in Fig.~\ref{f:pmu}.
There we display the pressure
$P$ as function of the baryon chemical potential $\mu_B$ for
baryonic and quark matter phases.
In the upper panel we show the results obtained using $V_1=0$,
whereas in the lower panel calculations with $V_1=0.01$ GeV (according to the
indication of the constraint in Eq. (\ref{constraint}) ) are displayed.
The solid line represents the calculations performed with the BBG method with
nucleons, and the other lines represent results obtained with different
choices of the gluon condensate $G_2$. We recall that the chosen values of
$G_2$ give values of the critical temperature in a range between $160$ and
$190$ MeV \cite{sim22}.

We observe that the crossing point is significantly affected by the value
of the gluon condensate, and only slightly by the chosen value of the
potential $V_1$. Moreover, with increasing $G_2$, the onset of the phase
transition is shifted to larger chemical potentials. Hence, we expect that
the neutron star will possess a thicker hadronic mantle with increasing $G_2$.

\begin{figure}[t] %............................................................
\centering
\includegraphics[width=12cm,angle=270]{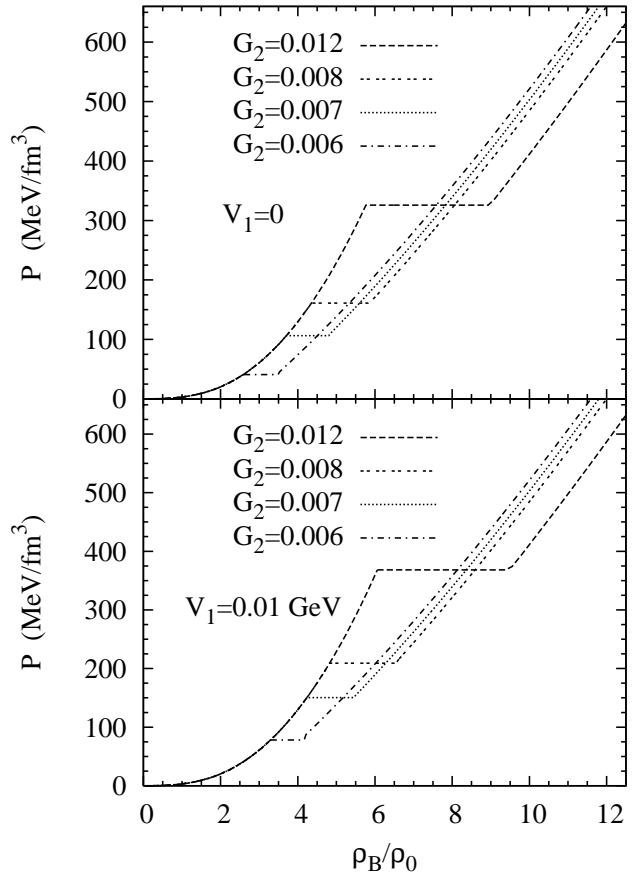}
\caption{
Pressure as a function of the baryon density, normalized with respect to
the nuclear matter saturation density $\rho_0$.}
\label{f:prho}
\end{figure}

In Fig.~\ref{f:prho} we display the total EoS, i.e. the pressure as a function
of the
baryon density for the several cases discussed above. The plateaus are
consequence of the Maxwell construction. Below the plateau,
$\beta$-stable and charge neutral stellar matter is in the purely
hadronic phase, whereas for density above the ones characterizing the
plateau, the system is in the pure quark phase.

 %................................................................

\begin{figure*}[t] %.............................................
\centering
\includegraphics[width=6cm,angle=270]{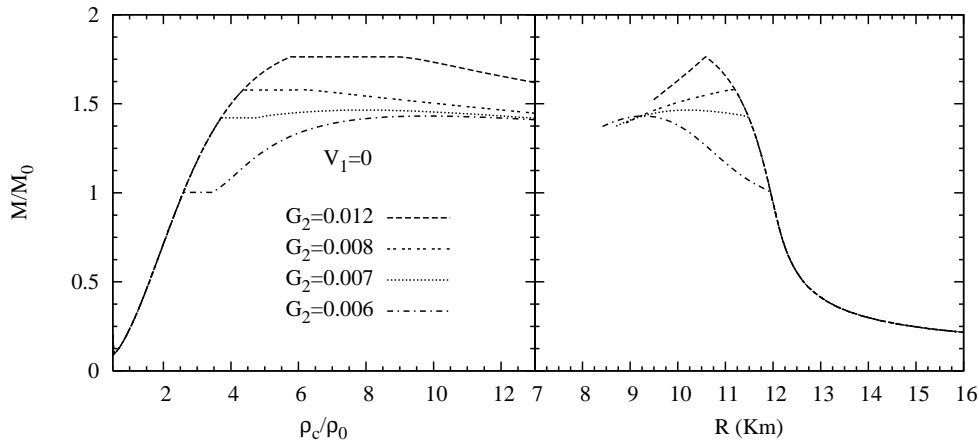}
\caption{
The gravitational mass (in units of the solar mass) is displayed as a function
of the central baryon density, normalized with respect to
the nuclear matter saturation density $\rho_0$ (left panel), and the
corresponding radius (right panel).}
\label{f:mrho_v0}
\end{figure*} %.......................................................

\begin{figure*}[t] %..........................................
\centering
\includegraphics[width=6cm,angle=270]{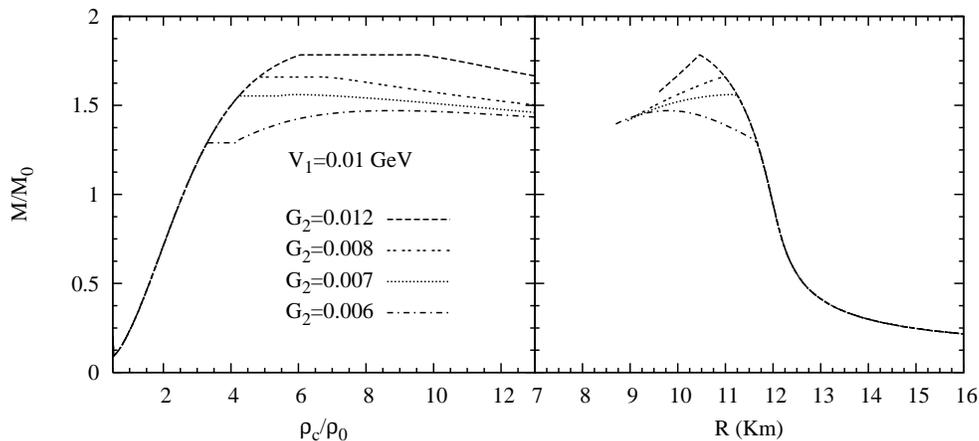}
\caption{Same as Fig.3 for $V_1=0.01~\rm GeV$.}
\label{f:mrho_v1}
\end{figure*} %................................................

The EoS is the fundamental input for solving the
well-known hydrostatic equilibrium equations of Tolman, Oppenheimer, and Volkov
\cite{shapiro}
for the pressure $P$ and the enclosed mass $m$
\bea
 {dP(r)\over dr} &=& -\frac{Gm(r)\epsilon(r)}{r^2}
 \frac{\big[ 1 + {P(r)\over\epsilon(r)} \big]
       \big[ 1 + {4\pi r^3P(r)\over m(r)} \big]}
 {1-{2Gm(r)\over r}} \:,
\label{tov1:eps}
\\
 \frac{dm(r)}{dr} &=& 4\pi r^{2}\epsilon(r) \:,
\label{tov2:eps}
\eea
being $\epsilon$ the total energy density ($G$ is the gravitational constant).
For a chosen central value of the energy density, the numerical integration of
Eqs.~(\ref{tov1:eps}) and (\ref{tov2:eps}) provides the mass-radius relation.
For the description of the neutron star crust,
we have joined the equations of state above described with the
ones by Negele \& Vautherin \cite{nv} in the medium-density regime
($0.001\;\mathrm{fm}^{-3}<\rho<0.08\;\mathrm{fm}^{-3}$),
and the ones by Feynman, Metropolis, \& Teller \cite{fmt}
and Baym, Pethick, \& Sutherland \cite{bps}
for the outer crust ($\rho<0.001\;\mathrm{fm}^{-3}$).

In Fig.\ref{f:mrho_v0} we display in the left panel the gravitational mass
(in units of solar mass $M_\odot = 2\times 10^{33}g$) as a
function of the central baryon density (normalized with respect to the saturation value),
and the corresponding
radius in the right panel. We observe that the value of the maximum mass spans over a range
between 1.4 and 1.8
solar masses, depending on the value of the gluon condensate $G_2$, as shown in Table I.
The stability of the pure quark phase appears
only for small values of $G_2$, which are hardly in agreement with observational data. In
fact, we recall that any
``good'' equation of state must give for the maximum mass at least 1.44 solar mass, the best
measured value so
far \cite{taylor}. By increasing the value of $G_2$, the maximum mass increases as well, up
to about 1.8 solar
mass, but the stability of the pure quark phase is lost, and the maximum mass contains in
its interior at most a
mixed quark-hadron phase. By switching on the potential $V_1$, as displayed in
Fig.\ref{f:mrho_v1}, we observe a trend similar to
the case $V_1=0$.
Therefore, generally speaking we can conclude that this model gives values of the maximum
mass in any case below
two solar mass, in agreement with the current observational data. However, the observational
data indicate that
NS with a mass of at least 1.6 solar masses do exist \cite{vela}, and this puts
a serious constraint on the value of the
gluon condensate, which is not easy to reconcile with the value 0.006 GeV$^4$, extracted
from the comparison
with the lattice data on the critical temperature. This result emphasize the relevance of
astrophysical data in
testing different quark matter models.

\begin{table} %......................................................
\begin{center}
\bigskip
\begin{ruledtabular}
\begin{tabular}{l|ccccc}
 $V_1$                 & $G_2$ & $M_G/M_\odot$ & $R$ (km) & $\rho_c/\rho_0$ \\
\hline
                       & 0.012      & 1.76      & 10.58  & 5.76  \\
 0.                    & 0.008      & 1.58      & 11.21  & 4.35  \\
                       & 0.007      & 1.46      & 10.2  & 7.92   \\
                       & 0.006      & 1.43      & 9.27  & 9.85  \\
\hline
                       & 0.012        & 1.78    & 10.46  & 6.06  \\
 0.01                  & 0.008        & 1.66    & 10.99  & 4.82  \\
                       & 0.007        & 1.55    & 11.26  & 4.23   \\
                       & 0.006        & 1.47    & 9.79   & 8.81   \\
\end{tabular}
\end{ruledtabular}
\end{center}
\caption{
Properties of the maximum mass configuration for different values
of the model parameters.}\label{t:mass}
\end{table} %..................................................................

\section{Discussion and Conclusions}

\begin{figure}[h] %............................................................
\centering
\includegraphics[width=6cm, angle=270]{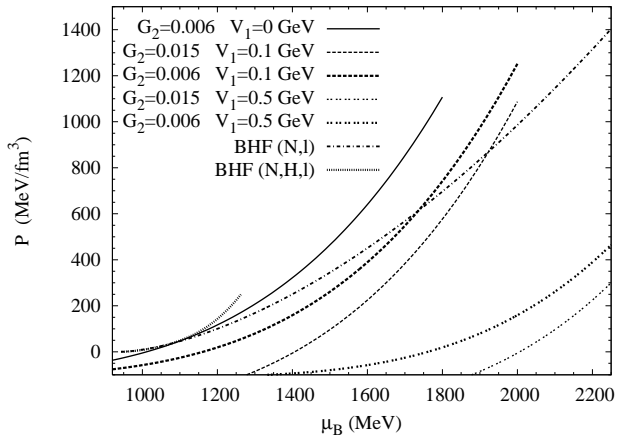}
\caption{Pressure as a function of the baryon chemical potential,
for different values of $V_1$ and $G_2$. The uppermost curve represents
the BHF EoS with hyperons.} \label{f:eosv1}
\end{figure}

The problem of the appearance of quark matter in the NS core has been discussed by
considering the microscopic
EoS in the FCM where the dynamics of confinement is assumed to be a long range
phenomenon. The
results confirm the idea
that confinement plays an important role to obtain a stable system under the gravitational
pressure. However, in this case, pure quark matter can appear only for a certain range of the gluon
condensate, which is mainly a parameter of the model.
The comparison with phenomenological data on NS masses gives strong constraints on
the values of this parameter,
which unfortunately are only marginally compatible with the range extracted by
comparing the model with lattice data at zero chemical potential.
However in this case the value of the large distance static $Q \bar Q$ potential $V_1$ turns
out to be very small. Other choices are possible if Eq. (\ref{d1nonpt}) is assumed to be valid only at long range,
while Eq. (\ref{norm}) is a true short range relationship. In this case the parameters $D_1^E(0)$ in the two
equations cannot be identified and may correspond to two different numerical
values, and therefore the value of $V_1$ must be considered an independent parameter. In the comparison
with lattice  calculations \cite{sim4} one finds a value $V_1\sim 0.5$ GeV at the critical temperature and for
$\mu = 0$. Besides that, the assumption of the independence of $V_1$ on $\mu$ can be questionable, it appears in
any case that the value of this parameter at high density and low temperature is quite uncertain. We have
therefore varied the strength of $V_1$ from  the small values previously considered up to $0.5$ GeV. The results
for the EOS is reported in Fig. \ref{f:eosv1} for different values of $V_1$. One can see that the hadron-quark
phase transition is shifted to higher values of the chemical potentials and therefore of the density. This can
be expected just by inspection of the formula for the pressure, which is clearly a decreasing function of $V_1$.
Actually already for $V_1 = 100$ MeV the phase transition cannot occur in NS, which is then composed of baryon
matter only, with a maximum mass around 2 solar masses. For higher values of $V_1$ the transition can possibly
occur only at exceedingly high values of the density, and therefore the quark phase is irrelevant for NS
physics.

 %.....

These results indicate once more a direct link between the NS quark content and the properties of deconfinement
in the hadron-quark phase transition. More quantitatively, if one considers that the well established values of
NS masses never exceed $\approx 1.6$ solar masses, then these observational data constrain  $V_1$ to small
values and in a narrow range, well below 100 MeV, in sharp contrast with values around $0.5$ GeV extracted from
lattice calculations. Despite the FCM is in good agreement with full QCD lattice data and is a well defined
theoretical approach where confinement is, ab initio, the crucial dynamical aspect, some refinements seem to be
needed once the astrophysical data are considered.

A relevant point to be clarified is the possible presence of hyperons,
whose onset is expected to be around 2-3 times the saturation density.
In Fig.\ref{f:eosv1} we plot two curves corresponding to the BHF EoS with
and without the inclusion of hyperons. As displayed,
both curves coincide at small values of the baryon chemical potential and,
after the onset of the hyperons, the former curve grows faster becoming the
uppermost one in the figure.

Therefore, only at  small values of $V_1$, of the order of 0.01 GeV or below,
the transition to quark matter occurs at about the same density as the hyperon
onset, as displayed in Fig. \ref{f:prho} and \ref{f:eosv1}. On the other 
hand, Fig.\ref{f:eosv1} clearly shows that, when increasing $V_1$ up to $V_1
\approx 0.5$ GeV, no crossing with the quark matter EOS is possible. We remark that the baryonic EoS with
hyperons in the BHF framework produces a maximum mass close to 1.3 $M_\odot$, below the observational limit, and
therefore it is not acceptable \cite{vid06}.

It has to be pointed out that in all cases where no phase transition to quark matter is
possible, with or
without hyperons, nuclear matter can reach densities where baryons are so closely packed
that keeping their
identity is highly questionable. This is the main physical qualitative argument that
suggests as likely a
transition to quark matter.

Another approximation used in the FCM is the so called Single Line Approximation
where the relevant dynamics is related to the interaction of a single quark or
gluon with the vacuum. At large density this could be no longer true, but in
the FCM the most important non perturbative effects are included in the field
correlators and in particular in the gluon condensate which drives the
transition. Therefore at large density the main effect should be related
to the $\mu$ dependence of $G_2$ in Eq. (\ref{norm}).
Lattice data at large temperature and small density show that the Color
Electric
condensate goes to zero at the transition point and the Color Magnetic condensate
survives at large temperature. In our analysis the same behavior has been
assumed at small temperature and large density (see Eqs. (\ref{pqgp1}) and (\ref{pqgp2})).
Of course our results
depend on this assumption and  we checked if a different qualitative
conclusion is reached with  a density dependent gluon condensate.
Following the suggestion in \cite{balcasza} for the density dependence of $G_2$
we obtain the same qualitative results with the possibility of a larger
NS mass $\simeq 2 $ solar masses.

\begin{acknowledgments}
The authors warmly thank Y. Simonov for enlightening discussions, and the
critical reading of the manuscript.
\end{acknowledgments}

\end{document}